\documentstyle[preprint,aps]{revtex}
\begin{document}
\draft
\tighten

\preprint{IC/96/76}

%----------------------------------------------------------------------
\title{Broken Symmetries in the Early Universe\thanks{ Based on
 talks by G.S. given at  {\em
Future Perspectives in Elementary Particle Physics} (Valencia, June 1995),
{\em Four Seas Conference} (Trieste, July 1995) and {\em 5th Hellenic School
 on Elementary Particle Physics} (Corfu, September 1995)  }}

%----------------------------------------------------------------------

\author{ \bf Alejandra Melfo}
\address{{\em SISSA, 34014 Trieste, Italy}, and 
{\em CAT, Universidad de Los Andes, M\'erida, Venezuela}}

\author{\bf Goran Senjanovi\'c} 
\address{\em ICTP, 34100 Trieste, Italy}
\maketitle       
%----------------------------------------------------------------------
\begin{abstract}
The year is 10$^{10}$ B.C. All the symmetries of Nature broken 
at low temperatures are completely restored. All of them?
 \newline No! \newline
 A tiny space of parameters, near the nonperturbative region, is there 
to resist
now and ever to the invading forces of symmetry restoration.  And life is not
easy for the thermally produced strings, monopoles and domain walls...
  
\end{abstract}

%%%%%%%%%%%%%%%%%%%%%%%%%%%%%%%%%%%%%%%%%%%%%%%%%%%%%%%%%%%%%%%%%%%%%%%%
\section{Introduction}\label{introduction}

Both common sense and daily life experience suggest the existence of phase 
transitions in systems exposed to temperature changes, leaving one with 
 the belief that a hotter environment normally implies more symmetry. And
 yet there are counterexamples, such as the Rochelle salt, which actually
 exhibit contrary behavior. It has been known now for some time that in
 particle physics systems, at least in theories beyond the standard model, 
the question of symmetry patterns at high temperature is rather complex, and
 depends on the parameter space of the theory under discussion.  It is our 
aim here to provide a short (and still somewhat pedagogical) review of this
 phenomenon.

Our motivation is at least twofold. The main reason for this study is simply
 curiosity; one wants to know if the complicated systems of present-day 
particle physics mimic more familiar ones, such as water or a ferromagnet.
 However, our interest is not purely academic. The physics of the standard
 model  of electro-weak interactions is based on the idea of spontaneous 
symmetry breaking, and thus what happens at high T could in principle help 
to probe the nature of this mechanism. Now, the relevant temperatures are
 too high to be of direct laboratory significance, but on the other hand the
 early universe can serve as an ideal place to study this important issue. 
The cosmological implications of high temperature symmetry behavior are
 profound and have to deal with such central questions as baryogenesis, 
the monopole and domain wall problems, the dynamics of cosmic strings, etc. 

Before addressing this issue in detail, we wish to say a few words about the
 early work in the field. The original work by Kirzhnitz and Kirzhnitz and
 Linde \cite{kl72}, suggested that at sufficiently high temperatures
 spontaneously broken symmetries are restored in a phase transition. This
 conclusion has been strengthened in the classic papers of Weinberg \cite{w74}
and Dolan and Jackiw \cite{dj74}, but remarkably enough, already then Weinberg
 notes the possibility of symmetry nonrestoration at high T in theories with 
more than one Higgs multiplet (he cites Coleman as being behind this 
observation). Interestingly, this went unnoticed for some years until the
 work of Mohapatra and Senjanovi\'c \cite{ms79}. They were  mainly motivated
 by the question of spontaneous CP violation at high temperatures for the
 sake of baryogenesis, and found out that it was possible to keep CP broken 
in a multi-Higgs $SU(2)\times U(1)$ model (recall that the idea of
 spontaneous CP violation requires necessarily more than one $SU(2)
 \times U(1)$ Higgs doublet). Much to their surprise, symmetry nonrestoration,
 as we have said, had already been discussed by Weinberg.

In \cite{ms79} it was also pointed out that symmetry nonrestoration may 
provide a way out of the domain wall problem, without fully addressing the
 question though. Soon after them Langacker and Pi \cite{lp80} pointed out 
that the same 
phenomenon may provide a way out of the monopole problem if electromagnetic 
gauge $U(1)$ invariance were to be broken in the early universe. However, 
the examples provided by all the above were  in some sense 
{\em ad hoc}, since they 
were achieved by enlarging the the minimal models just to serve this purpose.

Recently, the issue of high T symmetry behavior was readdressed in a series
 of papers \cite{ds95,dms95,dms96} devoted to the minimal, already accepted 
particle physics models with special emphasis on the domain wall and monopole
 problems. Here we review the central results of this study.
In the next section, after a general discussion on symmetry nonrestoration, 
we give some examples on how it can be realized in the context of global and 
local symmetries, paying particular attention to some interesting examples: 
spontaneous CP violation, Peccei-Quinn symmetry, and $SU(5)$ GUT. 
Then in section
 \ref{topdef}, we show how this is related to topological defects production 
and how it can lead to a solution of the monopole and domain wall problems in
 those theories. The final section contains a brief outlook.

\vspace{2cm}
\section{Broken symmetries at high temperature}
\label{broken}

The issue of what happens to a spontaneously broken symmetry when temperature
 effects are taken into account was addressed many years ago   
\cite{kl72,w74,dj74}. When the temperature reaches 
values much bigger than the Higgs field mass, its effects can be accounted 
for (up to the one-loop level) by a mass term in the effective potential
 proportional to $T^2$. More precisely, for a general Higgs potential
 written in terms of N real fields $\varphi_i$, the temperature contribution
 for $T >> m_\varphi$ is

\begin{equation}
 \Delta V(T) = {T^2 \over 24} \left[ \left({\partial^2 V \over \partial
\varphi_i \partial\varphi^i} \right)  +
 3 (T_a T_a)_{ij} \, \varphi^i \varphi^j \right]
\label{wf}
\end{equation}

where sum over repeated indices is assumed. This term being positive, 
it would unavoidably imply that a critical temperature 
will be reached above which the mass term for the Higgs is positive, 
restoring the symmetry. While this is certainly true for theories involving 
only one Higgs,  when two or more
 fields are responsible for the symmetry breaking, this need not be the
 case \cite{w74,ms79}. Consider for example a simple such theory, with a
$U(1)\times U(1)$ global 
symmetry and two complex Higgs fields $\phi$ and $\chi$ and a potential 

\begin{equation}
V = -{m^2_\phi \over 2} \phi^*\phi + {\lambda_\phi \over 4} (\phi^*\phi)^2
 - {m^2_\chi \over 2} \chi^*\chi + {\lambda_\chi \over 4} (\chi^*\chi)^2 + 
{\alpha \over 2}  \phi^*\phi \chi^*\chi
\label{simpot}
\end{equation}

and calculate the effective masses at high temperature using (\ref{wf})

\begin{eqnarray}
m^2_\phi(T) &= -m^2_\phi + {T^2 \over 12} ( 2\lambda_\phi + \alpha) 
\equiv -m^2_\phi
+ T^2 \nu_\phi^2 \nonumber \\
m^2_\chi(T) &= -m^2_\chi + {T^2 \over 12} ( 2\lambda_\chi + \alpha) 
\equiv -m^2_\phi
+ T^2 \nu_\chi^2 
  \label{simmass}
\end{eqnarray}

The crucial point is that the coupling constant alpha enters the mass
 terms at high temperature. Nothing forces $\alpha $ to be positive,
 all that is required is that the potential (\ref{simpot}) is bounded 
from below, which implies

\begin{equation}
\lambda_\phi \lambda_\chi > \alpha^2
\label{bound}
\end{equation}

One can have $\alpha <0$, and for example  $\lambda_\phi > 2 |\alpha| > 4 
\lambda_\chi$. Then $\nu_\chi$ in (\ref{simmass}) is negative, and $m_\chi(T)$ 
is negative for all temperatures. Notice that (\ref{bound}) prevents us from
 taking {\em both} $\nu_\chi$ and $\nu_\phi$ negative. Then one of the the 
$U(1)$ groups is  broken for any 
value of T. 
One can conceive of a model in which there is only one $U(1)$ 
symmetry, by including the term

\begin{equation}
\beta_1 \phi^*\chi \chi^*\chi + \beta_2 \chi^*\phi \phi^*\phi + h.c.
\end{equation}

in the potential. In this case, the cubic terms will force both of the vev's 
to be nonzero, even if only one of the masses is negative. The $U(1)$ symmetry 
 is completely broken at high temperature, for the same range of parameters
 as before.

The question of restoration becomes then a {\em dynamical} one, 
depending on the parameters of the potential.

\subsection{Global symmetries: $O(N_1) \times O(N_2)$ model}
\label{global}

It is not difficult to convince oneself that nonrestoration of symmetries 
is also possible in more complicated and realistic theories.  Suppose that 
the  fields in the previous example transform under more complicated groups.
 One would have then a bigger variety of possible self-couplings and 
couplings with the other field, introducing a number of coupling constants.
 The conditions of boundedness of the potential analogous to (\ref{bound})
can be many, and very complicated. However, it is enough that nonrestoration 
occurs for a reasonable range of parameters, so it is perfectly natural to
 ask for some of the couplings to be  small. Then one can consider only 
those couplings analogous to the ones of the simple model. That is, for
 fields ($ \Phi$, $\Xi$ ) transforming under  the representations $R_1$, $R_2$ 
of some group
 $G$ containing $N_1$, $N_2$ real fields ($\phi$, $\chi$), write the 
Higgs potential as

\begin{equation}
V = \sum_{a=1}^{N_1} \sum_{b=1}^{N_2}\left\{-{m^2_\phi \over 2} \phi^a\phi_a 
+ {\lambda_\phi \over 4} (\phi^a\phi_a)^2 
 - {m^2_\chi \over 2} \chi^b\chi_b + {\lambda_\chi \over 4}
 (\chi^b\chi_b)^2 - 
{\alpha \over 2}  \phi^a\phi_a \chi^b\chi_b \right\} + V_s
\label{onpot}
\end{equation} 

where $V_s$ contains terms whose coupling constants are assumed to be much 
 smaller than $\lambda_\phi$, $\lambda_\chi $ and $\alpha $. Thus in this case
the symmetry is $O(N_1) \times O(N_2)$. We will use the $O(N_1) \times O(N_2)$
models as a prototype that can effectively mimic more complicated groups.

 Taking $\alpha<0$,
the condition for the boundedness of the potential is again (\ref{bound}). 
The high temperature contributions to the masses are

\begin{eqnarray}
\Delta m_\phi^2(T) &= T^2 \nu_\phi^2  = T^2 \left[\lambda_\phi
 \left({2 + N_1 \over 12} \right) - {N_2 \over 12} \alpha \right] \nonumber \\
\Delta m_\chi^2(T) &= T^2 \nu_\chi^2  = T^2 \left[\lambda_\chi
 \left({2 + N_2 \over 12} \right) - {N_1 \over 12} \alpha \right]
\label{onmass}
\end{eqnarray}

and the $G$ symmetry will not be restored if the couplings lie in the range 

\begin{equation}
\lambda_\phi > \left({2 + N_2 \over N_1} \right) \alpha > 
\left({2 + N_2 \over N_1} \right)^2 \lambda_\chi
\label{onrange}
\end{equation}

Some relevant features of the range (\ref{onrange}) are worth mentioning.
\begin{itemize} 
\item notice that there is no lower bound on the smallest coupling, so
 one can always take it small enough to avoid the danger of the  
couplings getting too large and in conflict with perturbation theory.
This is not the case if $G$ is a gauge symmetry, since the gauge coupling
 will have to enter in the discussion, as we will see later. 

\item the conditions are weaker if  the ratio $N_2/N_1$ is big, that is, 
it will be easier for the representation with fewer real fields to maintain
 its vev at high temperature. 

\end{itemize}

In the simple example considered, only one of the fields can have a vev. 
This means that any subgroup of $G$ preserved by its vev will be restored. 
But condition (\ref{bound}) only prevents us from taking both mass terms
negative, and with an adequate coupling one can have both vevs nonzero even
 if one of the masses is positive. 
This will be the case if for example the symmetry allows for terms of the 
type $\phi^3 \chi$, as we saw in the general example. It is possible then 
to keep $G$ completely broken. 
We will illustrate how the mechanism can actually work by considering two 
examples: a discrete symmetry (CP) and a global U(1) symmetry (Peccei-Quinn).

\subsubsection{Spontaneous CP violation}
\label{cp}

Generally speaking, models of spontaneous symmetry breaking cannot be 
analized as suggested before, by taking some of the couplings to be
 negligible and considering only two self-couplings and a mixed one. 
For instance, in T.D. Lee's \cite{l73}
 original model of spontaneous CP violation, 
the CP-violating phase is the relative phase between the vevs of two 
doublet fields $\Phi_1, \Phi_2$, and it  appears due to the presence of
 terms of the type $\Phi_1^\dagger \Phi_2 \Phi_2^\dagger \Phi_2$. To
 have CP nonrestored, it is not enough to keep the vev's nonzero, one has 
also to make sure that the phase persists. In \cite{dms96} it has been
 shown that neither the T. D. Lee model, nor the model of CP violation
 with three doublets \cite{w76}, allow for nonrestoration of CP.

There is however one model of spontaneous CP violation which in addition 
has the nice feature of providing natural flavor conservation, where 
nonrestoration is easily achieved \cite{bb90,dms96}. It is a minimal
 extension of the 
Standard Model with the addition of a singlet field ($S$), odd under
 CP, and an additional down quark, with both left and right components
 $D_L^a$ and $ D_R^a$.  
 singlets under $SU(2)$. 

The interaction Lagrangian for the down quarks, symmetric under CP, 
contains the terms

\begin{eqnarray}
{\cal L}_Y &=& (\bar{u} \bar{d} )^a_L h_{a} \Phi D_R + 
(\bar{u} \bar{d} )^a_L h_{ab} \Phi d^b_{R} \nonumber \\
& &  + M_D \bar{D_L} D_R + M_a( \bar{D_L} d^a_{R} + h.c.) \nonumber \\
& & + i f_D S (\bar{D_L} D_R - \bar{D_R} D_L)
+ i f_a S (\bar{D_L} d^a_{R} - \bar{d^a_{R}}D_L )
\label{sinyuk}
\end{eqnarray}

Clearly, when $S$ gets a vev (at a scale $\sigma$ above the 
weak scale $M_W$) CP is  spontaneously broken by the terms in the last line.
CP violation at low energies is then achieved by complex phases appearing 
in the CKM  matrix through the mixings of $d$ and $D$ quarks. 

 The most general potential for the fields $\Phi$ and $S$  can be written as

\begin{eqnarray}
V(\Phi, S) &= & - m_\Phi^2 \Phi^\dagger \Phi +
 \lambda_\Phi (\Phi^\dagger \Phi)^2 \nonumber \\
& & - {m_S^2 \over 2} S^2 + {\lambda_S \over 4}  S^4 - 
{\alpha \over 2}   \Phi^\dagger \Phi S^2
\label{sinpot}
\end{eqnarray}

and it has a minimum at

\begin{equation}
\langle \Phi \rangle = {1 \over \sqrt{2}}
\left(\begin{array}{c} 0\\v \end{array} \right)
\;\;\; ; \;\;\; \langle S \rangle =  \sigma
\end{equation}

We can use here the general equations of the previous section, (\ref{bound}), 
(\ref{onmass}) and  (\ref{onrange}) with 
$N_1= 4$ and $N_2 = 1$. Notice that although the high-T mass of the doublets
 will contain the gauge coupling, it will not appear in the conditions upon 
the coupling constants, since we will only require that the mass of the 
singlet is negative at high T. This can be achieved if the couplings fall
 in the range

\begin{equation}
\lambda_\Phi > {3 \over 2} \alpha > 
\left(3 \over 2 \right)^2 \lambda_S
\label{sinrange}
\end{equation}

Thus CP can be violated at all temperatures.

\subsubsection{Peccei-Quinn symmetry}
\label{pq}

A very illustrative example of nonrestoration of a physically relevant
 symmetry is that of the $U(1)_{PQ}$ global symmetry, whose spontaneous 
breakdown provides a solution to the strong CP problem \cite{pq77}.
  In the invisible
 axion version \cite{k79} of the Peccei-Quinn mechanism, $U(1)_{PQ}$ is 
broken down at
 a scale $M_{PQ}$ much bigger than the QCD scale by the vev of a singlet
 field. The model requires in addition two doublet fields ($\phi_1, \phi_2$)
 that will couple
 to the quarks. Under $U(1)_{PQ}$ they  transform as

\begin{equation}
\phi_1 \rightarrow e^{i \alpha }\phi_1 \;\; ; \; \; \; \phi_2 \rightarrow 
e^{- i \alpha }\phi_2 \;\; ; \; \; \; S \rightarrow e^{2 i \alpha }S
\label{pqsim}
\end{equation}

The Higgs potential, invariant under $SU(2)_L \times U(1)_Y \times U(1)_{PQ}$ 
is written

\begin{eqnarray} 
V_{PQ} &=& \sum_i{\left[-{m_i^2 \over 2}  \phi_i^\dagger \phi_i + 
{\lambda_i \over 4} (\phi_i^\dagger \phi_i)^2\right]} 
-  {\alpha \over 2}(\phi_1^\dagger \phi_1)(\phi_2^\dagger \phi_2)-
 {\beta \over 2}(\phi_1^\dagger \phi_2)(\phi_2^\dagger \phi_1) \nonumber \\
&-& {m_s^2 \over 2} S^* S + {\lambda_s \over 4} (S^* S)^2 
- \sum_i{({\gamma_i \over 2} \phi_i^\dagger \phi_i}) S^* S
 - M (\phi_1^\dagger \phi_2 S + \phi_2^\dagger \phi_1 S^*) 
\label{pqpot}
 \end{eqnarray}

For $\beta >0$, the minimum is found at

\begin{equation}
\langle \Phi_i \rangle =\left(\begin{array}{c}
0 \\v_i \end{array}\right)  \;\; \; ; \;\;\;
 \langle S \rangle = v_S
\label{pqvev}
\end{equation}

To have $U(1)_{PQ}$ broken at any temperature, it is enough to keep the vev
 of the singlet nonzero for all T, and as before, we can use the formulas for
 the global case.  The high temperature mass for $S$ is
\begin{equation}
m_S^2(T) = - m_S^2 + {T^2 \over 3} (\lambda_S -\gamma_1 -\gamma_2)
\label{pqmass}
\end{equation}

But since we have three fields in this model, 
conditions (\ref{bound}) have to be generalized. 
Taking $v_S \gg v_i$, they are,
 to leading order

\begin{equation}
\lambda_i >0 \;\;\;, \;\; \lambda_S>0 \;\;\; ;\;\; 
\lambda_i\lambda_S>\gamma_i^2 \;\;\; ;\;\;
 \lambda_1\lambda_2 > (\alpha +\beta)^2 
\label{pqcond1}
\end{equation}
\begin{eqnarray}
& &Mv_s^3\left[{v_1^3 \over v_2}(\lambda_1\lambda_S-\gamma_1^2) +
{v_2^3 \over v_1}(\lambda_2\lambda_S-\gamma_2^2) 
- 2 v_1v_2 (\lambda_S(\alpha+\beta) + \gamma_1 \gamma_2)\right] \nonumber \\
&+ & v_S^2 v_1^2 v_2^2 \left[\lambda_1\lambda_2\lambda_S - 
\lambda_1\gamma_2^2 - \lambda_2\gamma_1^2  -\lambda_S (\alpha+\beta)^2
- 2 \gamma_1\gamma_2(\alpha+\beta)\right] >0
\label{pqcond2}
\end{eqnarray}

It is easily proven that it is possible to require
 $\gamma_1 +\gamma_2 > \lambda_S$, thus having a negative mass for $S$, 
without contradicting conditions (\ref{pqcond1}) and (\ref{pqcond2}). 
One can then have $U(1)_{PQ}$ broken at arbitrarily 
high temperatures.

\subsection{Gauged case}

As we have already mentioned, when the symmetry is gauged nonrestoration 
is not  straightforward. The gauge coupling provides a lower 
bound on the
 coupling constants, and depending on the particular gauge group chosen, 
one can then have to require the coupling constants to be of order one, 
away from the perturbative regime. 

To see it explicitly, consider a simplified model as the one of section 
\ref{global}, that is one where only the relevant  coupling constants 
are taken into account, and now the group $G$ is gauged.  
The two fields $\Phi$ and $\Xi$ transform under the
 representations  $R_i$ ($i= 1, 2$) whose generators satisfy

\begin{equation}
Tr(T_i^a T_i^b) = c_i \delta^{ab}
\end{equation}

Then the high-temperature masses are

\begin{eqnarray}
\Delta m_\phi^2(T) &= T^2 \nu_\phi^2  = T^2 \left[\lambda_\phi
 \left({2 + N_1 \over 12} \right) - {N_2 \over 12} \alpha + {1 \over 4} g^2 
{Dim(G) \over N_1}  r_1 c_1 \right] \nonumber \\
\Delta m_\chi^2(T) &= T^2 \nu_\chi^2  = T^2 \left[\lambda_\chi
 \left({2 + N_2 \over 12} \right) - {N_1 \over 12} \alpha + {1 \over 4} g^2 
{Dim(G) \over N_2} r_2 c_2 \right]
  \label{gaugemass}
\end{eqnarray}

where $g$ is the gauge coupling, $Dim(G)$ is the dimension of the group
and $r_i$ is 1 when the representation contains real fields, 2 when it is 
complex.
Asking $\nu_\chi$ to be negative and at the same time the fulfillment of
 the bound (\ref{bound}) now implies

\begin{equation}
\lambda_\phi > {\alpha^2 \over \lambda_\chi} >
{1 \over \lambda_\chi}\left[ \left(N_2 + 2 \over N_1 \right) \lambda_1
 + 3 g^2 {Dim(G) \over N_1 N_2} r_2 c_2  \right]^2
\end{equation}

As a function of $\lambda_\chi$, $\lambda_\phi$ has a minimum at

\begin{equation}
\lambda_\chi = 3 g^2 {Dim(G)\over N_2 (2 + N_2)} r_2 c_2
\end{equation}

So  $\lambda_\phi$ is bounded from below as

\begin{equation}
\lambda_\phi > 12 g^2 { (2 + N_2) Dim(G) r_2 c_2 \over N_1^2 N_2} 
\label{gaugerange}
\end{equation}

The dimension of the representation $R_1$ (under which the fields that 
looses its vev transforms) now plays an even  more fundamental 
role: it has to be big enough, if we want perturbation theory to be valid. 
This is better illustrated by a concrete example

\subsubsection{SU(5) and nonrestoration}

Being the simplest of GUTs, it is only natural to investigate the high 
temperature behavior of $SU(5)$. The usual pattern of symmetry breaking 
goes trough the Standard Model, as

\begin{center}
$SU(5) \stackrel{\langle H \rangle}{\rightarrow} SU(3)_C\times SU(2)_L 
\times U(1)_Y  \stackrel{\langle \Phi \rangle}{\rightarrow}
 SU(3)_C\times U(1)_{em}$
\end{center}

where $H$ is taken to transform under the adjoint representation, while 
$\Phi$ can be either in the 5-dimensional fundamental representation or,
 if one requires a realistic theory of fermion masses, in the 45-dimensional
 representation. 

In ref. \cite{kst90}, a range of parameters was considered for which $\Phi$ 
(in the 5-dimensional representation) keeps its vev at high T, thus 
preventing the restoration of $SU(2)_L 
\times U(1)_Y$ of the Standard Model. Here we consider the case in which $H$ 
keeps its vev, a case that may have interesting cosmological consequences 
\cite{dms95}, and that will be particularly illustrative.

First suppose that $\Phi$ is in the five-dimensional representation. Then in
(\ref{gaugerange}), setting $N_1=10$, $N_2=24$,  $c_2 = 5$ $r_2=1$, we
get

\begin{equation}
\lambda_\phi >  {78 \over 5} g^2 
\end{equation}

For a typical value of $g^2 \sim 1/4$, we find $\lambda_\phi/4$ dangerously
close to one.
On the other hand, if we take the more realistic model where $\Phi$ is in 
the 45-dimensional representation, we have $N_1=90$, and then the lower limit
is $9^2$ smaller

\begin{equation}
\lambda_\phi >  {26 \over 135} g^2 
\end{equation}

So it will be safe to take $\lambda_\phi \sim 0.05$. It is then possible to
 have $SU(5)$ broken at high temperatures.

\section{Topological Defects}
\label{topdef}

Symmetry nonrestoration can be used in certain theories to cure the problems 
related to topological defects. Topological defects such as monopoles,
 strings and domain walls, can arise in cosmological phase transitions, which
 are a direct consequence of symmetry restoration at high temperature. Namely,
 if symmetries are restored by thermal effects, one has a picture in which
 they become  broken as the universe cools down. The fact that the Higgs
 field is only causally correlated inside a finite region at a given time,
 then, gives rise to defects via the so-called Kibble mechanism \cite{k76} 

Of the three kinds of defects mentioned, only cosmic strings are compatible
 with the standard cosmology. Monopoles are produced in a phase transition
in too big numbers \cite{p79}, and domain walls are too heavy
\cite{zko74}: in both cases the result 
is that they overclose the universe.

As was suggested in \cite{sss85,ds95,dms95}, one way out could be to 
avoid the phase transition. In theories with more than one Higgs fields, 
this can be done in principle by requiring that the parameters of the
 potential fall into the ranges where nonrestoration, if possible,  can occur.

The theories exhibiting  nonrestoration considered in the previous section, 
are of the kind that admit topological defects. The theory of CP violation,
based on a spontaneously broken discrete symmetry, has domain wall solutions. 
The global U(1) symmetry of Peccei-Quinn allows for the formation of global 
 strings, however when the QCD scale is reached,  the Nambu-Goldston boson 
associated with its breaking (the axion) acquires a vev. When this happens, 
the strings become the edges of domain walls, which are stable. Finally,  
when $SU(5)$ breaks down to $SU(3)\times SU(2) \times U(1)$, monopoles are 
produced. As we have seen, there is a natural way to avoid the restoration 
of the symmetries at high temperature, i.e., to avoid the phase transition.
 Defects are then not produced via the Kibble mechanism.

However, it is still true that the theory admits the classical solutions
 that we call defects. In order not to actually have these structures formed,
we have to make sure, to start with, that 
``initially'', i.e. at the Planck scale, the field is distributed uniformly
 over scales that are not causally correlated, at least over a scale of the
 size of the comoving horizon. This is the same as requiring that the 
so-called horizon problem be solved, for example by invoking an era of
 primordial inflation. But even if this condition is satisfied, thermal 
fluctuations can drive the field away from the minimum chosen. So one has to 
take into account the possibility of thermal production of  defects, as we 
do now.

\subsection{Domain Wall problem}

Consider the nucleation of a large spherically symmetric domain wall or
 a closed loop of string. The production  rate per unit time per unit volume
at a temperature $T$ will be given by \cite{l81}

\begin{equation}
\Gamma = T^4 \left({S_3\over 2\pi T }\right)^{3/2} e^{-S_3/T}
\label{rate}
\end{equation}

where $S_3$ is the energy of the closed defect. The suppression factor
$e^{-S_3/T}$ is readily calculated in the limit where the defect's radius
is much bigger than its width.  For the domain walls produced in the model  
of CP violation with a singlet, we get

\begin{equation}
{S_3 \over T} \gg {16 \pi \over 3 \sqrt{6}} {\sqrt{2\alpha - 3\lambda_S}
\over \lambda_S}
\label{sup1}  
\end{equation}

 Analogously, for the Peccei-Quinn model the thermal production of
large loops of strings is suppressed by 

\begin{equation}
{S_3 \over T} \gg
4 \pi^2 {\sqrt{\gamma_1 + \gamma_2 - \lambda_S} \over \lambda_S}
\label{sup2}
\end{equation}
 
We see that in both cases, it suffices to take the singlet's
 self-coupling $\lambda_S$ small to avoid significant thermal production  
 of defects.

\subsection{Monopole problem}
 Monopoles  can be thermally produced in $e^+ e^-$   (and other
 charged particles) collisions. Turner \cite{t82} has investigated the
 conditions under which the density of thermally produced monopoles will
 be consistent with cosmology, and found that we should have

\begin{equation}
{m_M \over T} \geq 35
\label{lim}
\end{equation}

where $m_M$ is the monopole mass. More precisely, for $m_M/T \geq 20$, 
he that

\begin{equation}
{n_M \over n_\gamma} \simeq 3\times 10^3 \left( m_H \over T \right)^3 
e^{- 2 m_M/T}
\label{elim}
\end{equation}

where $n_\gamma$ is the photon density; and from the upper limit
 $n_M/ n_\gamma\leq 10^{-24}$, one obtains (\ref{lim})

Now, in $SU(5)$  the lightest monopoles weigh \cite{dt80}
 
\begin{equation}
m_M = {10 \pi \over \sqrt{2} g} v_H
\end{equation}

For $g^2/ (4 \pi) \simeq 1/50$ or $g \simeq 1/2$, $m_M \simeq 40 v_H$,
 and thus the consistency with the cosmological bound (\ref{lim}) implies

\begin{equation}
{v_H \over T} \geq 1
\label{vt}
\end{equation}

Obviously  this will put even more restrictive conditions on the
 parameters of the theory. For the simplified $O(N_1) \times O(N_2)$  models 
we have considered (with $N_2 = 24$ and $N_1 = 10 $ or $90$ , we have 
at high temperature

\begin{equation}
{v_H^2 \over T^2} = - {\nu_H^2 \over \lambda_H} >1
\end{equation}

instead of just $\nu_H^2 > 0$. Condition (\ref{gaugerange}) becomes now

\begin{equation}
\lambda_\phi > 12 g^2 { (N_2 + 14) Dim(G) c_2 \over N_1^2 N_2} 
\label{monorange}
\end{equation}

We have for the case in which $\Phi$ is in the 45-dimensional representation

\begin{equation}
\lambda_\phi > { 38\over 135} g^2 
\end{equation}

which is perfectly compatible with perturbation theory still. We conclude that 
thermally produced monopoles can be kept below the density limit required 
by cosmology.

\vspace{2cm}

\section{Outlook} 

  We have illustrated how spontaneously
 broken symmetries may and may not be restored at high temperature. We have 
also  suggested that symmetry nonrestoration may provide a way out of 
the domain wall 
and monopole problems. The latter may even be cured in the canonical $SU(5)$
theory, especially if one accepts the necessity of a \underline{45} of Higgs
 to reproduce correctly the quark and lepton masses.

Our discussion so far has been based only on the leading one-loop computation
 of high temperature scalar masses. The situation becomes more complicated
 when the next-to-leading effects are included, as recently pointed out by
 Bimonte and Lozano \cite{bl95} and Roos \cite{r95}. Bimonte and Lozano even 
find out that the $SU(5)$ example discussed above may be in trouble; 
more precisely that one may be taken out of the perturbative regime. We feel
 that more study is needed before one has a conclusive answer on these
 issues, but we should add that if they are right, one would be forced 
to turn one's attention to more complicated (and possibly more realistic) 
theories such as $SO(10)$, characterized by multistage symmetry breaking
 patterns. The work on this is now in progress. 

We have also left out the supersymmetric theories. Here unfortunately we have
 a no-go theorem   due to Mangano and Haber \cite{m84} which states that
 internal symmetries in the context of SUSY are necessarily restored at 
high T. As we were preparing this for print, a paper of Dvali and Tamvakis
 \cite{dt96} has appeared which tries to offer a possible way out using 
higher dimensional nonrenormalizable interaction.

\vspace{2cm}

\section*{Acknowledgments}

G.S. would like to acknowledge the original collaboration with Rabi 
Mohapatra and both of us the collaboration with Gia Dvali. 

%%%%%%%%%%%%%%%%%%%%%%%%%%%%%%%%%%%%%%%%%%%%%%%%%%%%%%%%%%%%%%%%%%%%%%%%

%%%%%%%%%%%%%%%%%%%%%%%%%%%%%%%%%%%%%%%%%%%%%%%%%%%%%%%%%%%%%%%%%%%%%%%%
\end{document}